\documentclass[conference]{IEEEtran}
\IEEEoverridecommandlockouts

\usepackage{cite}
\usepackage{hyperref}
\usepackage{url}
\usepackage{booktabs}
\usepackage{amsfonts}
\usepackage{amsmath}
\usepackage{nicefrac}
\usepackage{microtype}
\usepackage{xcolor}
\usepackage{graphicx}
\usepackage{float}
\usepackage{comment}
\usepackage{multirow}
\usepackage{booktabs}

\title{Scaling LLM-Driven Multi-Agent Systems:\\
Design Principles and Architectural \\ Scalability Analysis}

\author{
Linus Sander\textsuperscript{1,*},
Fengjunjie Pan\textsuperscript{1,*},
Vahid Zolfaghari\textsuperscript{1},
Andre Schamschurko\textsuperscript{1},
Nenad Petrovic\textsuperscript{1},
and Alois Knoll\textsuperscript{1}%
\thanks{\textsuperscript{*}Linus Sander and Fengjunjie Pan contributed equally to this work.}%
\thanks{\textsuperscript{1}The authors are with Robotics, Artificial Intelligence and Real-Time Systems,
School of Computation, Information and Technology, Technical University of Munich,
Munich, Germany.
\{linus.sander, f.pan, v.zolfaghari, andre.schamschurko,
nenad.petrovic, k\}@tum.de.}%
}

\begin{document}
\maketitle

% -----------------------------------------------------------------
\begin{abstract}
% -----------------------------------------------------------------
LLM based multi-agent systems have the potential to enable collective intelligence and scale toward solving highly complex tasks through coordinated ensembles of specialized agents. However, despite their theoretical potential, the architectural design space remains largely non-systematized and lacks broadly established design principles. Furthermore, the scalability characteristics of such systems are only partially understood so far.
This paper makes two contributions. We first distill four design principles for scalable MAS architectures from a structured analysis of prior work: simplicity, elastic feedback, sequential workflows with optional loops, and summary-based communication. We operationalize these principles in a reference architecture whose topology is formalized as a constrained directed workflow graph, and we evaluate four configurations of increasing complexity on a standardized benchmark of terminal-based system engineering tasks using two LLMs of differing capability. 
Our findings show that scaling yields measurable accuracy improvements with approximately linear cost growth, but only when the underlying LLM exceeds a minimum capability threshold. Performance peaks at intermediate complexity, then degrades due to timeouts and evaluation limitations. In addition, persistent consistency issues emerge as a central challenge across all scaling levels. These results provide concrete design guidance for practitioners and highlight consistency and evaluation standardization as key targets for future research.

\end{abstract}

% =================================================================
\section{Introduction}
\label{sec:introduction}
% =================================================================

Building intelligent agent systems has been a longstanding goal in AI research~\cite{russell1998learning,jennings1998roadmap}, and the rapid maturation of large language models (LLMs) has marked a significant step toward achieving it~\cite{xi2025rise}. The world knowledge embedded in LLMs~\cite{yu2023kola}, advanced reasoning and planning capabilities~\cite{achiam2023gpt} and strong generalizability~\cite{brown2020language} make them powerful foundations for autonomous decision-making and qualify them as natural core controllers in modern agent systems~\cite{wang2025survey}. Augmented with memory, sensors, and actuators, LLM based agents represent a promising path toward truly autonomous systems capable of handling complex, uncertain environments~\cite{yao2022react,masterman2024landscape}.
 
However, a single agent remains fundamentally bounded: its working memory is constrained by the LLM's context window~\cite{hsieh2024ruler}, its single perspective reduces robustness~\cite{he2025llm}, and sequential operation imposes a ceiling on task following~\cite{liu2023agentbench} and parallelism~\cite{koh2026multi}. These constraints, together with the rich communication capabilities of modern LLMs~\cite{chen2023agentverse} and insights from collective intelligence (CI) and human collaboration, motivate the shift to multi-agent systems (MAS)~\cite{guo2024large}. The principle is simple: multiple specialized agents communicate with each other and reason collectively toward a shared objective~\cite{li2023camel}.

 \begin{figure}[t]
\centering
\includegraphics[width=0.95\linewidth]{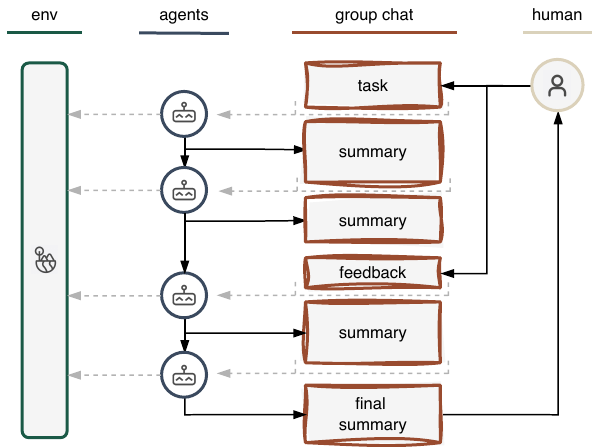}
\caption{Example of a multi-agent system lifecycle.}
\label{fig:figure_model_spoox_flow}
\end{figure}

Yet, designing an effective and robust MAS that is applicable to real-world domains remains a substantial engineering challenge~\cite{cemri2025multi}. The design space is vast, spanning communication topologies, agent specialization, workflow orchestration, and memory architectures~\cite{li2024survey}, and no consensus has emerged on consistent approaches to constructing robust and generalizable systems~\cite{han2024llm}. Moreover, scaling agent systems to increase capability inevitably grows complexity, giving rise to new failure modes: communication overload~\cite{yan2025beyond}, inter-agent inconsistency~\cite{becker2025stay}, loss of efficiency~\cite{cemri2025multi}, and loss of system-wide task alignment~\cite{han2024llm}. Prior scaling studies examine individual dimensions such as agent count~\cite{hong2023metagpt,qian2024scaling} or debate rounds~\cite{chan2023chateval} and consistently report bounded, architecture-dependent benefits, yet a systematic characterization of how accuracy, cost, and reliability jointly evolve under principled architectural scaling is still missing. 

This paper makes four contributions. First, we derive four literature-grounded design principles for scalable MAS architectures. Second, we present a reference architecture that operationalizes these principles through summary-based group communication and a formally constrained directed workflow graph with agent-controlled feedback loops. Third, we report an empirical scalability analysis across four configurations and two model tiers, characterizing accuracy, cost, consistency, and feedback loop utilization. Fourth, we distill our findings into actionable design guidance and identify run-to-run consistency as the central open problem for production-grade MAS.
 
%The remainder of this paper is organized as follows. Section~\ref{sec:background} reviews the foundations of LLM based agents and multi-agent systems and positions our study within prior work on MAS for software engineering and MAS scaling. Section~\ref{sec:principles} derives the design principles, and Section~\ref{sec:architecture} presents the reference architecture. Section~\ref{sec:setup} describes the experimental design, and Section~\ref{sec:results} reports the results. Section~\ref{sec:discussion} discusses the findings and their implications, and Section~\ref{sec:validity} examines threats to validity. Sections~\ref{sec:future} and~\ref{sec:conclusion} outline future directions and conclude.
 
\section{Background and Related Work}
\label{sec:background}
% =================================================================

\subsection{LLM-Based Multi-Agent Systems}
\label{sec:mas_arch}

LLM-based agents operate through an iterative observation--reasoning--action loop and are commonly augmented with role instructions, tools, memory, and feedback mechanisms~\cite{yao2022react,wang2024survey,shinn2023reflexion}. Multi-agent systems (MAS) extend this architecture by distributing responsibilities across agents that coordinate toward a shared objective~\cite{guo2024large,talebirad2023multi}.

Two architectural dimensions are particularly relevant~\cite{li2024survey}. \emph{Agent profiles} define roles, responsibilities, tool access, and positions in the workflow~\cite{li2023camel,hong2023metagpt}, while the \emph{communication strategy} determines which agents interact, what information they exchange, and how control moves through the system~\cite{yan2025beyond}. Common structures include sequential, hierarchical, debate-based, and group communication topologies, with coordination mechanisms such as peer feedback, debate, and leader orchestration~\cite{du2023improving,masterman2024landscape}. However, additional interaction can introduce redundant information and coordination overhead~\cite{chan2023chateval}. MAS performance therefore depends not only on agent count, but also on how roles and information flows are organized for the task~\cite{han2024llm}.

\subsection{Multi-Agent Systems for Software Engineering}

Software engineering (SE) is well suited to agent-based collaboration because tasks can be decomposed across roles and evaluated through executable feedback. Tests, runtime errors, and static-analysis results provide signals for iterative repair, making testing and review common components of SE-oriented MAS~\cite{chen2023teaching,xia2025demystifying,huang2023agentcoder}.

Existing systems range from role-rich workflows, such as \emph{MetaGPT}, \emph{ChatDev}, and \emph{AgileCoder}~\cite{hong2023metagpt,qian2024chatdev,nguyen2025agilecoder}, to more minimal approaches, including \emph{AutoCodeRover}, \emph{Agentless}, and \emph{SWE-agent}~\cite{xia2025demystifying,yang2024swe}. The latter show that strong search, editing, and execution capabilities can rival more elaborate agent organizations. Terminal-based coding agents further extend this setting to interactive system-engineering tasks in which agents inspect unfamiliar environments, execute tools, interpret feedback, and revise their plans~\cite{claude_code_git,openai_codex_git}. This tool-rich setting forms the evaluation domain of our study.

\subsection{Scaling Behavior of Multi-Agent Systems}
\label{sec:scaling_related}

MAS can be scaled through agent count, communication rounds, feedback-loop depth, and repeated planning or reflection~\cite{tillmann2025literature}. Prior work reports improvements from additional agents or coordination~\cite{hong2023metagpt,qian2024scaling}, but these gains are often bounded. In debate-based systems, performance may plateau or decline as communication becomes redundant and context becomes saturated~\cite{chan2023chateval,liu2024groupdebate}. Dynamic agent generation and repeated repair workflows similarly show that scaling benefits depend on the underlying architecture and task~\cite{perera2025auto,huang2023agentcoder,yan2025beyond}.

Most prior studies vary a single scaling dimension, focus on static reasoning benchmarks, and report average task success. It therefore remains unclear how accuracy, cost, communication load, and run-to-run consistency jointly change as a coherent MAS architecture is scaled in an interactive, tool-rich environment. We address this gap by evaluating progressively more complex configurations within a shared reference architecture.

% =================================================================
\section{Design Principles}
\label{sec:principles}
% =================================================================
Four design principles, drawing on the survey findings presented above, form a foundation for scalable MAS architectures.

\textbf{P1 (Simplicity):} MAS design often yields unnecessarily complex systems, a tendency considered a poor design pattern. It undermines transparency, coordination efficiency, real-world applicability, and rigorous evaluation~\cite{cemri2025multi,nguyen2025agilecoder,kapoor2024ai}. Minimalistic approaches challenge this tendency directly: designs such as Agentless and the Mini SWE-Agent~\cite{minisweagent_git} demonstrate that simplicity is fully compatible with strong performance. Architectural complexity should therefore be introduced incrementally and only where empirical need has been established.

\textbf{P2 (Elastic feedback):} Feedback, in all its forms, is one of the most fundamental mechanisms for enhancing MAS performance~\cite{guo2024embodied,chen2023agentverse,masterman2024landscape}. Effective systems must be able to adapt to increasing task complexity without sacrificing efficiency on simpler tasks~\cite{hou2024large,wang2024survey}, and defining principled termination conditions remains an open challenge~\cite{he2025llm}.
In MAS debate architectures, additional debate rounds tend to help only up to a point, with the optimum depending on architecture and task~\cite{becker2025stay,tillmann2025literature}. Similarly, practitioners developing SE-domain MASs highlight the importance of designing flexible feedback cycles~\cite{nguyen2025agilecoder,huang2023agentcoder,qian2024chatdev}.
Ultimately, the optimal number of feedback cycles is task-dependent and cannot be predetermined. Architectures should let individual agents assess at runtime whether further feedback is warranted, dynamically balancing refinement with forward progress.

\textbf{P3 (Sequential workflow with optional loops):} Scaling collaborating agents requires careful orchestration~\cite{guo2024large} and a communication strategy that sustains rising volume~\cite{he2025llm,masterman2024landscape,yan2025beyond}. Furthermore, the system's topology and workflow must dynamically accommodate tasks of varying complexity while avoiding overload and interaction chaos ~\cite{cemri2025multi,guo2024embodied}. Yet, overly dynamic and flexible topologies, may undermine consistency and robustness, and can drive up token costs~\cite{talebirad2023multi,tillmann2025literature}.
Focusing on a specific domain helps clarify the task structure, environment, useful agent roles, and typical workflow phases and feedback loops. This approach has proven effective in the SE domain~\cite{nguyen2025agilecoder,qian2024chatdev}. 
We argue that, given a clear understanding of the target domain, one can construct a primarily sequential workflow with selectively inserted loops back to earlier stages. This yields a topology that balances stability and adaptability: the core workflow stays fixed, while feedback loops add flexibility where needed. The shortest path handles simple tasks efficiently and loops accommodate complex cases.

\textbf{P4 (Summary-based communication):} Scaling an MAS increases agents, interactions, and communication volume, making information overload and LLM context limits a real concern. During its turn, an agent reasons, uses tools, and iterates internally. None of these detailed action traces or full reasoning chains should be shared with subsequent agents. Instead, once the agent completes its work, it should share only a concise natural-language summary with the next agents to come. This preserves communication flexibility while suppressing noise propagation.

% =================================================================
\section{Reference Architecture}
\label{sec:architecture}
% =================================================================

The proposed framework operationalizes P1--P4 through two core components: a shared communication channel of agent summaries and a set of topology design rules.

\subsection{Group Chat of Summaries}

A shared group chat channel serves as the central repository for all inter-agent information. Upon completing its sub-task, each agent contributes a natural-language summary to this channel. The instruction is minimal by design, agents are simply told to write a summary of what they did when finished or stuck, granting expressive freedom through unconstrained natural language while mitigating unnecessary chatter. Subsequent agents receive the full accumulated group chat history as context, ensuring coherence and avoiding redundant work within a shared execution environment~\cite{han2024llm}. Since every agent produces one summary per step in the sequential workflow, the overall context grows linearly.

Figure~\ref{fig:figure_model_spoox_flow} illustrates the straightforward process: the chat is initialised with the user's task; the active agent reads all prior messages, performs its work, and appends a summary; the next agent then inherits this enriched context.

This design directly realises P1 and P4. The group chat also doubles as a concise progress overview and a natural entry point for human feedback: any message added by a user becomes part of the shared context and is automatically available to all subsequent agents. While summaries and shared message pools are well-established in MAS design~\cite{hong2023metagpt,liu2024groupdebate,nguyen2025agilecoder,qian2024chatdev}, our approach is distinguished by their combination with a sequential workflow and fully unfiltered context access.

\subsection{Directed Workflow with Optional Loops}

A well-designed workflow is crucial for MAS efficiency and for ensuring smooth collaboration among agents~\cite{cemri2025multi,han2024llm}. Based on principles P1--P3, we establish a set of general rules for topology design.

The topology is defined as a directed graph $G_w = (A, E, s, T)$, where $A$ is a non-empty set of individual agents (vertices), $E$ is a set of directed edges defining the workflow, $s \in A$ is the designated start agent (first invoked after task submission), and $T \subseteq A$, $|T| > 0$, is the set of terminal agents. Each edge connects two distinct agents, $\forall\,(u,v) \in E:\; u, v \in A \wedge u \neq v$, and the graph satisfies three structural constraints:
\begin{itemize}
\item Each non-terminal agent has at least one successor: $\forall\, u \in A \setminus T:\; |\{(u,v) \in E\}| \geq 1$.
\item Each non-start agent is reachable: $\forall\, v \in A \setminus \{s\}:\; |\{(u,v) \in E\}| \geq 1$.
\item Each agent has at most three successors: $\forall\, u \in A:\; |\{(u,v) \in E\}| \leq 3$.
\end{itemize}

These simple rules ensure a controlled yet flexible workflow and align naturally with the group chat strategy. However, one important mechansim remains to be specified: how and when is the next agent selected? This is delegated to the agents themselves. Each agent decides when its sub-task is done and which successor to activate.  Depending on the topology, this may involve a single fixed successor or a choice among several possible next agents. Candidates are described in the system prompt, and the agent signals its choice via a tag \verb+[<next_agent>]+ in its final summary. A parser detects the tag, posts the summary to the chat, and triggers the next agent.

We propose that combining a group chat communication mechanism, built on natural language summaries, with a sequential workflow enriched by optional loops forms a robust communication architecture well suited for scaling MASs and operating in complex environments.

The linearly growing group chat mitigates exploding context sizes, and the use of concise natural language summaries preserves expressiveness and maintains efficient information flow. Moreover, incorporating loops and allowing agents to locally decide whom to call next, enables workflows that adjust dynamically to progress and task difficulty. At the same time, the existence of a defined shortest path from the initial to the final agent ensures a minimal, reliable workflow, for maintaining quality and coherent collaboration.

% =================================================================
\section{Experimental Design}
\label{sec:setup}
% =================================================================

In this section, we describe the benchmark, agent specifications, models, system configurations, metrics, and procedure of the scalability analysis.

\subsection{Benchmark and Environment}

All configurations are evaluated on terminal-bench~\cite{merrill2026terminal}, a benchmark of more than 80 end-to-end system engineering tasks within a terminal environment. The tasks range from basic OS operations to complex server management workflows and SE challenges. Each task is executed within an isolated Docker environment, providing controlled and reproducible conditions backed by automated, verifiable succeass criteria. The domain requires multi-turn reasoning and interaction, as well as exploration of an initially unfamiliar environment, making it well-suited for meaningful agent evaluation~\cite{yao2022react,liu2023agentbench}.

An effective interface is fundamental to building a successful and practically useful agent, and primarily involves designing robust tools that equip the agent with straightforward and reliable capabilities~\cite{masterman2024landscape,ruan2023tptu,sager2025ai}. For this experimental study, four tools were developed:

\begin{itemize}
\item \textit{Shell} tool: executes a single Bash command in a non-persistent environment
\item \textit{Terminal} tool: maintains a persistent \textit{tmux} session, accepts arbitrary keystrokes, and returns a full 128$\times$16 terminal window
\item \textit{Python} tool: allows the definition and execution of complete Python scripts
\item \textit{Search} tool: based on LangChain's \textit{DuckDuckGoSearch}, queries the web and returns the top four results.
\end{itemize}

\subsection{Agent Roles and Prompt Design}

In our MASs, agent profiling is realised entirely through system prompt design. We drew inspiration from existing work in which full system prompts for MAS setups have been disclosed and analysed~\cite{chen2023teaching,hong2023metagpt,liu2023agentbench,qian2024chatdev}.

Each prompt was developed through an iterative cycle of drafting, LLM-assisted refinement, and evaluation via manual runs in a local Docker environment using simple test scripts. To ensure robustness against model variation~\cite{chatterjee2024posix,wang2024survey}, multiple LLMs were consulted across iterations, while also trying to keep prompts simple, generalisable, and focused on the target domain.

%This process nonetheless highlights the tedious nature of prompt design and the need for more structured techniques. While existing research offers useful guidance, the process still relies on iterative refinement, benchmark evaluation, and manual trace inspection to reduce misinterpretation from natural-language ambiguity~\cite{he2025llm,wang2024survey}.

All prompts follow a block-wise Markdown structure: a context block outlining the task environment; a task description block specifying the agent's role responsibilities and topology position; and an operational rules block comprising tool execution guidance, termination behaviour and additional operating rules motivated by established prompting concepts~\cite{kojima2022large,wei2022chain} % (TODO more)

%Table~\ref{tab:agents} provides an overview of all agent roles, toolsets, and scopes.

\begin{comment}
\begin{table*}[t]
\caption{Experimental design: agent roles, toolsets, and scopes.}
\label{tab:agents}
\centering
\begin{tabular}{llll}
\toprule
Agent            & Appears in         & Tools                         & Scope \\
\midrule
Singleton        & Singleton          & sh, terminal, py, search      & Full task \\
Explorer         & MAS-M, MAS-L       & sh, terminal, search          & Environment exploration only \\
Solver           & MAS-S, MAS-M       & sh, terminal, py, search      & Full solution implementation \\
Sub-Task Planner & MAS-L              & none                          & Planning only \\
Sub-Task Solver  & MAS-L              & sh, terminal, py, search      & One sub-task implementation \\
Tester           & MAS-S, MAS-M, MAS-L& sh, terminal, py              & Test \& report \\
Refiner          & MAS-L              & sh, terminal, py, search      & Bug fixing \\
Approver         & MAS-M, MAS-L       & none                          & Quality gate \\
Summarizer       & MAS-S, MAS-M, MAS-L& none                          & Final summary \\
\bottomrule
\end{tabular}
\end{table*}
\end{comment}

\subsection{Models}

We evaluate three LLMs of differing scale from the same model family: GPT-5-nano and GPT-5-mini as smaller models, and GPT-5.3-Codex as a state-of-the-art model. Holding the model family constant isolates the effect of base model capability from vendor-specific differences in behaviour. All three models are optimised for tool use~\cite{gpt-5-system-card} and are also used by other agents on the terminl-bench leaderboard. Moreover, the GPT model suite has proven effective across multiple MAS setups~\cite{guo2024embodied} and has demonstrated strong code generation quality~\cite{chen2023teaching}.

\subsection{Scaling Configurations}

\begin{figure}[t]
\centering
\includegraphics[width=0.75\linewidth]{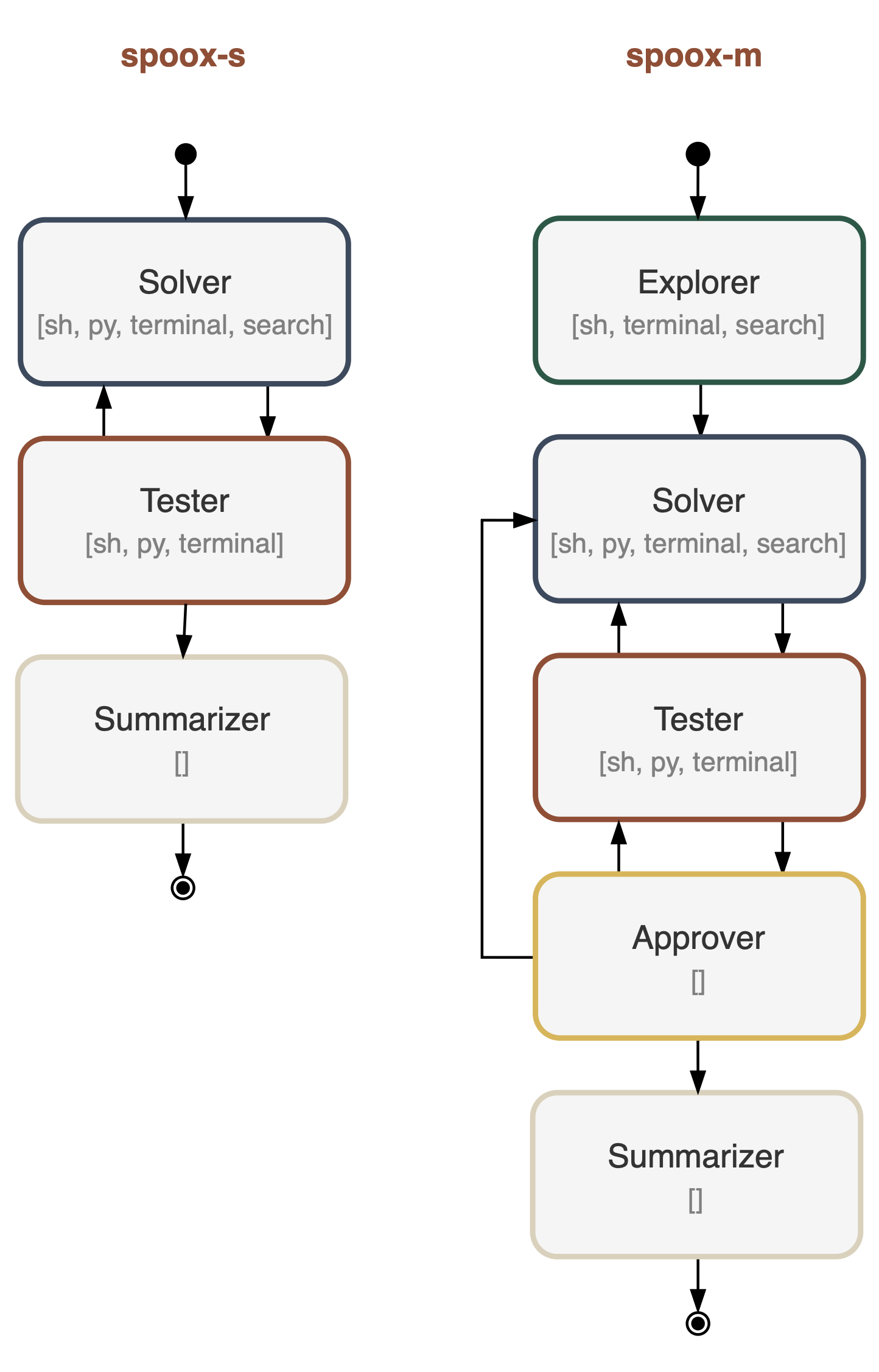}
\caption{Setups of reference MAS systems}
\label{fig:mas-examples}
\end{figure}

To support the scaling study's objectives, four MAS topologies of progressively increasing architectural complexity were instantiated following the proposed reference architecture. The central aim is to examine how growing levels of task decomposition and feedback-loop complexity influence collective reasoning dynamics in a MAS. The configurations scale incrementally from a single-agent baseline, with each step adding two agents and two feedback loops. Figure~\ref{fig:mas-examples} shows the MAS-S and MAS-M configurations as examples.

\textbf{Singleton} (1 agent, 0 loops): A single agent uses all available tools to solve the terminal task end-to-end, with no group chat or topology needed. As a well-designed MAS is expected to outperform single-agent systems~\cite{chen2023agentverse,chan2023chateval}, it serves as our  single-agent baseline.

\textbf{MAS-S} (3 agents, 1 loop): The Solver handles the full task, and the Tester evaluates the result, deciding whether to return control to the Solver for revision or pass the solution to the Summarizer. This loop exemplifies a guided yet elastic workflow, repeated as often as the Tester seems necessary.

\textbf{MAS-M} (5 agents, 3 loops): An Explorer conducts structured environment investigation before implementation begins, providing the Solver with an informed starting state, inspired by the localization and retrieval phases of minimalistic repository agents~\cite{xia2025demystifying,yang2024swe}. The Solver--Tester pairing from MAS-S is retained. An Approver agent makes a final holistic assessment downstream, with loop options back to both Solver and Tester. This adds quality assurance against inadequate verification and early termination~\cite{cemri2025multi,he2025llm}, mitigates Tester bias~\cite{wang2024large}.

\textbf{MAS-L} (7 agents, 5 loops): The solving phase is decomposed into a Sub-Task Planner and a Sub-Task Solver, separating planning from execution. The Planner may loop back to the Explorer for additional information gathering if needed. Following implementation, the system enters a Tester--Refiner--Approver loop, mirroring MAS-M but applied exclusively to the refinement phase. This introduces the most extensive task decomposition among the configurations, potentially enabling deeper SE refinement~\cite{he2025llm}.

\subsection{Metrics}

When benchmarking agents, accuracy is typically the primary metric, as it directly measures the task success rate~\cite{sager2025ai}. However, focusing solely on accuracy tends to produce overly complex and costly agents that are impractical for real-world deployment~\cite{kapoor2024ai}. Meaningful evaluation must therefore consider multiple cost aspects and behavioral metrics to obtain a realistic understanding of an agent system's efficiency and robustness~\cite{he2025llm,masterman2024landscape}. Additional metrics recorded:

\textbf{Test-weighted accuracy}: In terminal-bench, a task's reward is 0 if any of its unit tests fail. This hides important nuances of the agent's accuracy, like partial success when passing four of five tests. We therefore compute an additional accuracy metric as the fraction of passed unit tests per trial, and refer to the average across all tasks as \textit{test-weighted accuracy}.

\textbf{Costs}: In practical deployments, actual monetary costs and execution time become important factors~\cite{masterman2024landscape}. To capture a more meaningful performance profile, the costs analysis comprises: the number of LLM calls, prompt and completion tokens, as well as the systems's overall execution time.

\textbf{Consistency}: Running each agent–model configuration multiple times enables us to assess consistency. For every configuration, we can identify the tasks solved in at least one run. We then further divide this set into tasks solved by only one run, by two runs, and so forth. This breakdown offers insight into the reliability of the agents' performance.

\textbf{Loop utilization}: Evaluating the topology designs requires detailed inspection of actual agent execution sequences. In particular, we examine whether agents engage the elastic looping mechanisms at all. Loop usage is therefore tracked throughout evaluation.

\subsection{Procedure}

For GPT-5-mini and GPT-5-nano, each agent system pairing was evaluated across three to five terminal-bench runs. As MAS-M yielded the strongest results for both models, it was run five times to meet the requirements for official submission to the terminal-bench 2.0 leaderboard. To provide a state-of-the-art reference point, the MAS-M and GPT-5.3-Codex pairing was likewise evaluated over five runs.

% =================================================================
\section{Results}
\label{sec:results}
% =================================================================

In this section, we report the results of the scalability analysis, structured along the discussed metrics. Table~\ref{tab:results} summarizes the main outcomes.

\subsection{Accuracy}

\begin{figure*}[t]
\centering
\includegraphics[width=0.8\linewidth]{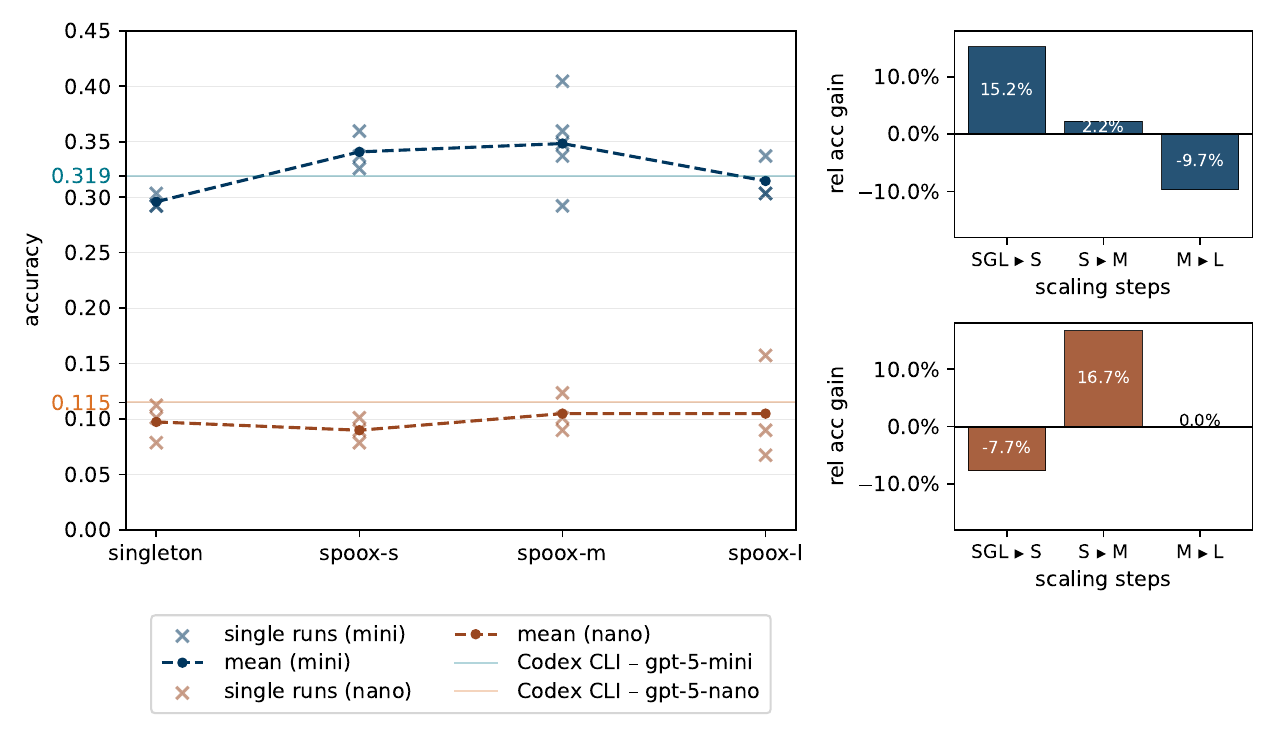}
\caption{Accuracy across agent-system scaling steps.}
\label{fig:accs}
\end{figure*}

Figure \ref{fig:accs} provides an aggregated view of all GPT-5-mini and GPT-5-nano runs, showing the accuracy computed as the proportion of successful out of the total number of trials. The result for every agent–model pairing is plotted, complemented by a dashed line to highlight the general mean accuracy trend. For reference, horizontal lines in matching colors indicate the performance of the leading agents on the official terminal-bench leaderboard, as retrieved on 21 July 2026. In addition, alongside the accuracy line plots, the relative accuracy gains for each scaling step are visualized as separate bar plots. 

For GPT-5-mini, accuracy improves across each scaling step from Singleton to MAS-M. The Singleton to MAS-S transition produced the largest single gain (15.2\% relative), reflecting the shift from a single to multi-agent system. MAS-M achieved peak accuracy, surpassing the top single-agent system on the public terminal-bench leaderboard. MAS-L, however, does not maintain the positive trend and showed accuracy below MAS-S, substantially driven by execution timeout violations. For GPT-5-nano, no consistent scaling trend appeared: accuracy remained nearly flat across all four configurations.

The \textit{test-weighted accuracy} reproduces the general GPT-5-mini scaling pattern, though less pronounced: the Singleton to MAS-S gain drops from 15.2\% to 3.0\%. MAS-M remains best ($\mu = 0.543$, $\sigma = 0.024$), followed by MAS-S ($\mu = 0.530$, $\sigma = 0.037$), MAS-L ($\mu  = 0.518, \sigma = 0.013$), and Singleton ($\mu  = 0.514$, $\sigma = 0.014$). For GPT-5-nano, performance rises from Singleton ($\mu = 0.328$, $\sigma = 0.072$) to a peak at MAS-M ($\mu = 0.351$, $\sigma = 0.029$) and drops to $\mu = 0.320$ ($\sigma = 0.114$) at MAS-L.

Additionally, the five MAS-M runs using GPT-5.3-codex yielded a mean accuracy of $\mu = 0.715$, placing the configuration in the middle of the field among all GPT-5.3-codex based agents on the official terminal-bench leaderboard. This demonstrates that the proposed architecture is competitive with other state-of-the-art agents.

\subsection{Consistency}

\begin{figure*}[t]
\centering
\includegraphics[width=0.8\linewidth]{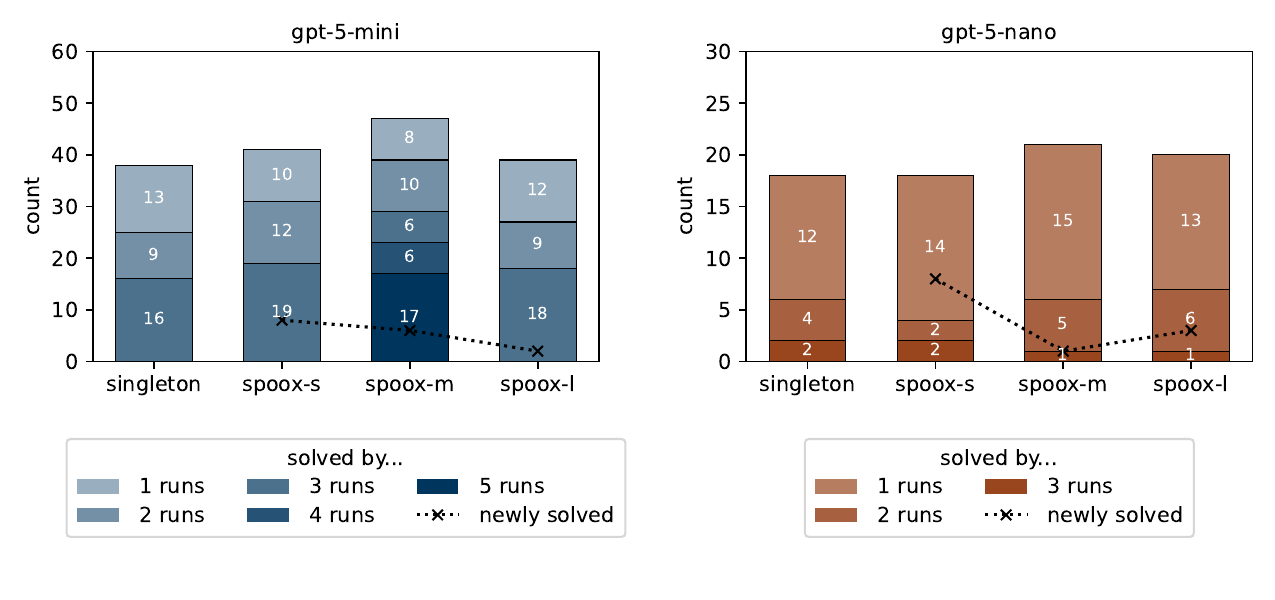}
\caption{Task-solving consistency across agent runs.}
\label{fig:accs-cons}
\end{figure*}

Figure~\ref{fig:accs-cons} evaluates task-solving consistency across runs. Bar height shows the number of tasks solved by at least one run per model–agent combination, and segments show how many runs solved them. The dotted black line visualizes the number of tasks solved by at least one run but not by any lower-ranked agent before. The term \textit{lower-ranked} refers to less scaled agents. For example, Singleton and MAS-S are considered lower-ranked than MAS-M.

In the GPT-5-mini experiments, the Singleton solved 38 of 89 unique tasks at least once across three runs. This increases up to MAS-M to 47 before dropping to 39 for MAS-L. Tasks solved by an agent that have never been solved by any lower-ranked agent, fall from 8 (MAS-S) to 6 (MAS-M) to 2 (MAS-L).  In addition, the segmented bars reveal that fewer than half of each agent's solved tasks were solved consistently across all runs.

For GPT-5-nano, consistency is notably lower: Singleton and MAS-S each solved two tasks across all runs, MAS-M and MAS-L only one. Moreover, the majority of tasks that were solved in just one of the three runs. Regarding tasks uniquely solved at each scaling level, MAS-S reached 8 tasks not solved by the Singleton, while MAS-M achieved 1 and MAS-L achieved 3.

\subsection{Costs}

Costs scaled approximately linearly with architectural complexity for both models. LLM call count, completion tokens, and execution time all increased steadily from Singleton to MAS-L. Beyond this, \textit{AgentTimeout} errors, triggered when terminal-bench interrupts an agent for exceeding the task timeout, rise with complexity for both models, from Singleton ($\mu = 3.33$ for GPT-5-mini and $\mu = 1.00$ for GPT-5-nano) to MAS-L ($\mu = 50.00$ for GPT-5-mini and $\mu = 37.67$ for GPT-5-nano), with the largest jumps at Singleton→MAS-S and MAS-M→MAS-L.

\subsection{Loop utilization}

In MAS-S, the single Tester--Solver loop is used more often with GPT-5-nano than GPT-5-mini: 71.9\% of GPT-5-mini trials (120/167) use no loop, versus 51.1\% (92/180) for GPT-5-nano.

MAS-M adds two more loops starting from the Approver and returning to the Tester or Solver. As with MAS-S, a larger share of GPT-5-mini trials uses no loop (180 of 257; 70.0\%), while the proportion is lower for GPT-5-nano (74 of 164; 45.1\%). The Tester--Solver loop remains the most frequently used, whereas the two Approver-based loops are engaged considerably less often (only in 19 GPT-5-mini and 14 GPT-5-nano task runs).

In MAS-L, nearly all trials use at least one loop, dominated by the SubTaskSolver--SubTaskPlanner loop. The SubTaskPlanner--Explorer return loop appears more often with GPT-5-nano, and the Refiner--Tester loop is triggered only in 20 (nano) and 17 (mini) times. The Approver loops are rare (once or twice per model), with the Approver--Tester loop never triggered for GPT-5-nano.

\begin{table}[t]
\caption{
Summary of scaling results. 
Cost trends are based on the average number of completion tokens.
Consistency (\%) is the share of solved tasks solved in \emph{all} runs.
}
\label{tab:results}
\centering
\begin{tabular}{llcccc}
\toprule
Model & System & Accuracy & Acc.\ $\Delta$ & Cost trend & Consistency \\
\midrule
\multirow{4}{*}{\rotatebox[origin=c]{90}{\shortstack{GPT-5\\mini}}}
  & Singleton & 0.296 & --                   & baseline     & 42\% \\
  & MAS-S     & 0.341 & $+$15.2\%            & $+$1$\times$ & 46\% \\
  & MAS-M     & 0.348 & $+$2.2\%            & $+$1.4$\times$ & 36\% \\
  & MAS-L     & 0.315 & $-$9.7\%             & $+$3.2$\times$ & 46\% \\
\cmidrule(lr){1-6}
\multirow{4}{*}{\rotatebox[origin=c]{90}{\shortstack{GPT-5\\nano}}}
  & Singleton & 0.097 & --                   & baseline     & 11\% \\
  & MAS-S     & 0.090 & $-$7.7\%             & $+$1.1$\times$ & 11\% \\
  & MAS-M     & 0.105 & $+$16.7\%            & $+$1.6$\times$ & 5\% \\
  & MAS-L     & 0.105 & $+$0.0\%             & $+$3.4$\times$ & 5\% \\
\bottomrule
\end{tabular}
% TODO(data): add standard deviations across runs per configuration if available.
\end{table}

% =================================================================
\section{Discussion}
\label{sec:discussion}
% =================================================================

In this section, we interpret the findings with respect to the research questions, relate them to documented MAS challenges, and derive implications for practice.

\paragraph{LLM capability as a prerequisite}
The most significant finding is that architectural sophistication provides no benefit, and may increase cost, when the underlying LLM is insufficiently capable. The GPT-5-nano results demonstrate that agents can engage the communication infrastructure, trigger feedback loops, and consume tokens without producing meaningfully better solutions. MAS architectures amplify the capabilities of capable LLMs but cannot substitute for them. This aligns with failure analyses that identify the selection of an inappropriate or insufficiently capable LLM as a system-level failure source~\cite{cemri2025multi}. Establishing base model adequacy before investing in architectural complexity is therefore the first practical design decision.

\paragraph{Peak performance at intermediate complexity}
MAS performance peaked at intermediate architectural complexity (MAS-M: 5 agents, 3 loops) rather than at the most complex configuration (MAS-L: 7 agents, 5 loops). This aligns with the logistic scaling behavior identified by Qian et al.~\cite{qian2024scaling} and peak-then-flatten patterns in debate-based MAS~\cite{du2023improving,liu2024groupdebate}. Potential reasons are multifaceted: longer minimum execution paths increase timeout exposure; additional agents introduce more error propagation opportunities~\cite{becker2025stay,masterman2024landscape}; and feedback loops provide no benefit if the agents responsible for triggering them choose not to do so.

\paragraph{Consistency as the central bottleneck}
The failure to improve run-to-run consistency across scaling configurations is the most practically consequential finding. Recent evidence confirms that MAS solutions are stochastic: the same agent system applied to the same task may succeed in one run and fail in another~\cite{mehta2026agentsdisagreethemselvesmeasuring}, and dedicated quality-assurance agents mitigate but do not eliminate this variability~\cite{venkataramani2026masprov}. Our results reproduce this pattern at every scaling level. A system that achieves higher mean accuracy on multi-run averages but cannot reliably solve individual tasks is of limited production value. The likely root cause is the tendency of downstream agents to accept and reinforce the framing established by upstream agents, mirroring the documented propagation of faulty responses between agents~\cite{guo2024large,masterman2024landscape}. When the Tester validates the Solver's interpretation, the Approver has little basis to override their mutual agreement, even if both have adopted a subtly incorrect framing. Breaking this would require better quality-assurance agents configured with explicit adversarial instructions: to probe, challenge, and invoke feedback loops proactively.

\begin{comment}
NOT INCLUDED FOR NOW -> NEED TO BE CHECKED IF BACKED BY SOURCES 
\paragraph{Relation to known MAS challenges}
Our findings empirically ground several challenges reported in the literature. The linear growth of the group chat confirms that summary-based communication can contain the information overload that otherwise grows with agent count~\cite{yan2025beyond}. The MAS-L degradation illustrates how coordination complexity and error propagation intensify with scale~\cite{cemri2025multi,han2024llm,becker2025stay}, and the low loop initiation by the Approver reflects the documented tendency toward premature validation and inadequate verification~\cite{cemri2025multi}. Finally, all configurations inherit LLM-level limitations, including hallucination~\cite{bang2023multitask,azamfirei2023large}, context degradation under long or noisy inputs~\cite{liu2024lost,shi2023large}, and prompt sensitivity~\cite{chatterjee2024posix,chen2023mapo}; our block-wise standardized prompts mitigate but cannot eliminate these effects.
\end{comment}

\paragraph{Implications for practice}
Our results translate into four practical guidelines: (1) establish LLM capability adequacy before investing in architectural complexity; (2) prefer simple architectures and expand them incrementally, monitoring the cost-accuracy tradeoff at each step; (3) design quality-assurance agents with adversarial rather than cooperative orientations; and (4) measure consistency alongside accuracy when evaluating MAS performance.

\section{Conclusion}
\label{sec:conclusion}
% =================================================================

In this paper, we investigated how architectural scaling affects the performance, cost, and reliability of LLM based multi-agent systems through a principle-guided empirical study. From a structured analysis of prior work on MAS design and scaling, we distilled four design principles for scalable architectures: simplicity (P1), elastic feedback (P2), sequential workflows with optional loops (P3), and summary-based communication (P4). We operationalized these principles in a reference architecture and instantiated four configurations of increasing complexity, from a single-agent baseline to a seven-agent system. All configurations were evaluated on a standardized terminal task benchmark using two LLMs of differing capability. Architectural scaling improved accuracy at approximately linear cost growth, but only above a minimum base model capability. Performance peaked at intermediate complexity, and run-to-run consistency remained below fifty percent of solved tasks across all configurations, identifying reliability rather than mean accuracy as the central obstacle to production-grade MAS. 

\begin{comment}
Our findings point to four directions for future research:

\paragraph{Adversarial quality assurance}
Improving consistency likely requires quality-assurance agents explicitly prompted to challenge rather than validate: probing robustness, seeking alternative framings, and invoking feedback loops proactively. This adversarial orientation departs from typical cooperative MAS designs but may be essential for production reliability.

\paragraph{Long-term memory}
Integrating long-term memory, through vector databases or structured interaction logs, would allow agents to accumulate domain knowledge across runs, shortening exploration phases and reducing repetitive reasoning~\cite{shinn2023reflexion}.

\paragraph{Model heterogeneity}
Assigning different models to different roles, for instance a capable model for Solver and Approver and a smaller model for less demanding agents, offers a path toward cost-efficiency without sacrificing accuracy at key decision points. This remains underexplored.

\paragraph{Evaluation standardization}
Progress in MAS research requires standardized multi-domain benchmarks assessing both accuracy and cost under realistic constraints, and consistency metrics that capture run-to-run reliability rather than averaged success rates.
\end{comment}

% =================================================================
\bibliographystyle{IEEEtran}   % todo check again which style required
\bibliography{main}
% =================================================================

\end{document}